\documentclass{article}
\usepackage[T1]{fontenc} 
\usepackage[utf8]{inputenc} 
\usepackage{ismir,amsmath,cite,url}
\usepackage{graphicx}
\usepackage{color}
\usepackage{amsfonts}
\usepackage{tabularx}
\usepackage{booktabs}
\usepackage{subcaption}
\usepackage{multirow,bigdelim}

\title{Passage summarization with recurrent models \\ 
for Audio -- Sheet Music Retrieval}

\multauthor
{Lu{\'i}s Carvalho$^1$ \hspace{1cm} Gerhard Widmer$^{1,2}$} { 
$^1$Institute of Computational Perception \& $^2$LIT Artificial 
Intelligence Lab\\
Johannes Kepler University Linz, Austria\\
{\tt\small \{luis.carvalho, gerhard.widmer\}@jku.at}
}

\def\authorname{L. Carvalho and G. Widmer}

\begin{document}

\maketitle
\begin{abstract}
Many applications of cross-modal music retrieval are related to connecting sheet
music images to audio recordings.
A typical and recent approach to this is to learn, via deep neural networks, a 
joint embedding space that correlates short fixed-size snippets of audio and 
sheet music by means of an appropriate similarity structure. 
However, two challenges that arise out of this strategy are the requirement of 
strongly aligned data to train the networks, and the inherent discrepancies of 
musical content between audio and sheet music snippets caused by local and global 
tempo differences.
In this paper, we address these two shortcomings by designing a cross-modal 
recurrent network that learns joint embeddings that can summarize longer passages 
of corresponding audio and sheet music. 
The benefits of our method are that it only requires weakly aligned audio -- sheet 
music pairs, as well as that the recurrent network handles the non-linearities 
caused by tempo variations between audio and sheet music. We conduct a number of 
experiments on synthetic and real piano data and scores, showing that our proposed 
recurrent method leads to more accurate retrieval in all possible configurations.
\end{abstract}
\section{Introduction}\label{sec:intro}

The abundance of music-related content in various digital formats, including 
studio and live audio recordings, scanned sheet music, and metadata, 
among others, calls for efficient technologies for cross-linking between documents
of different modalities.
In this work, we explore a cross-modal task referred to as audio -- sheet music passage 
retrieval. 
We define it as follows: given an audio fragment as a query, search within an image database and retrieve the corresponding sheet music passage; or vice versa, find 
the appropriate recording fragment given a query in the form of some snippet of (scanned) sheet music.
%

A fundamental step in audio–sheet music retrieval concerns defining a suitable 
shared representation that permits the comparison between items of different
modalities in a convenient and effective way.
The conventional approaches for linking audio recordings to their respective printed 
scores are based on handcrafted mid-level 
representations~\cite{MuellerABDW19_MusicRetrieval_IEEE-SPM, 
IzmirliS12_PrintedMusicAudio_ISMIR}.
These are usually pitch-class profiles, like chroma-based 
features~\cite{FremereyCME09_SheetMusicID_ISMIR, 
KurthMFCC07_AutomatedSynchronization_ISMIR}, symbolic 
fingerprints~\cite{ArztBW12_SymbolicFingerprint_ISMIR}, or the bootleg 
score~\cite{Tsai20_LinkingLakhtoIMSLP_Bootleg_ICASSP,
YangTJST19_Sheet2MIDIRetrieval_Bootleg_ISMIR}, which is a coarse
mid-level codification of the main note-heads in a sheet music image.
However extracting such representations requires a series of pre-processing 
stages that are prone to errors, for example optical music recognition on the 
sheet music side~\cite{Calvo-ZaragozaHP21_OMRReview_ACM, LopezVCC21_OMRAug_ICDAR,
EelcoU_OMRAugs_ISMIR}, and automatic music transcription on the audio 
part~\cite{BoeckS12_TranscriptionRecurrentNetwork_ICASSP, 
HawthorneESRSREOE18_OnsetsAndFramesPianoTrans_ISMIR, 
SigtiaBD16_DNNPolyPianoTrans_TASLP}.

A promising approach~\cite{DorferHAFW18_MSMD_TISMIR,
DorferAW17_AudioSheetCorrespondences_ISMIR} has been proposed to 
eliminate these problematic pre-processing steps by learning a shared 
low-dimensional embedding space directly from audio recordings 
and printed scores. 
This is achieved by optimizing a cross-modal convolutional network (CNN) to project
short snippets of audio and sheet music onto a latent space, in which 
the cosine distances between semantically related snippets are minimized, 
whereas non-related items of either modality are projected far from each other.
Then the retrieval procedure is reduced to simple nearest-neighbour 
search in the shared embedding space, which is a simple and fast algorithm.

\begin{figure}
 \centerline{
 \includegraphics[width=1\columnwidth]{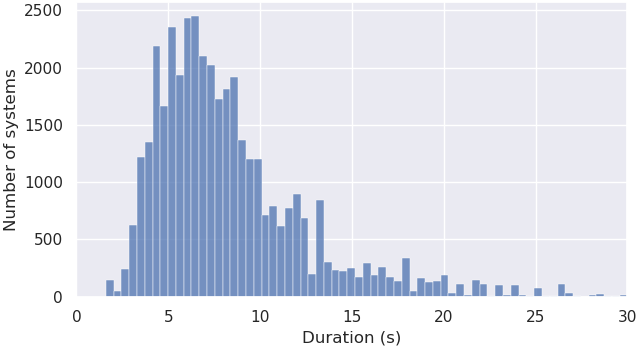}}
 \caption{Distribution of system durations in around 40,000 examples 
 from the MSMD. More than 25\% of the passages are longer than ten 
 seconds.}
 \label{fig:histogram}
\end{figure}

\begin{figure*}[t]
  \centering
  \includegraphics[width=1\textwidth]{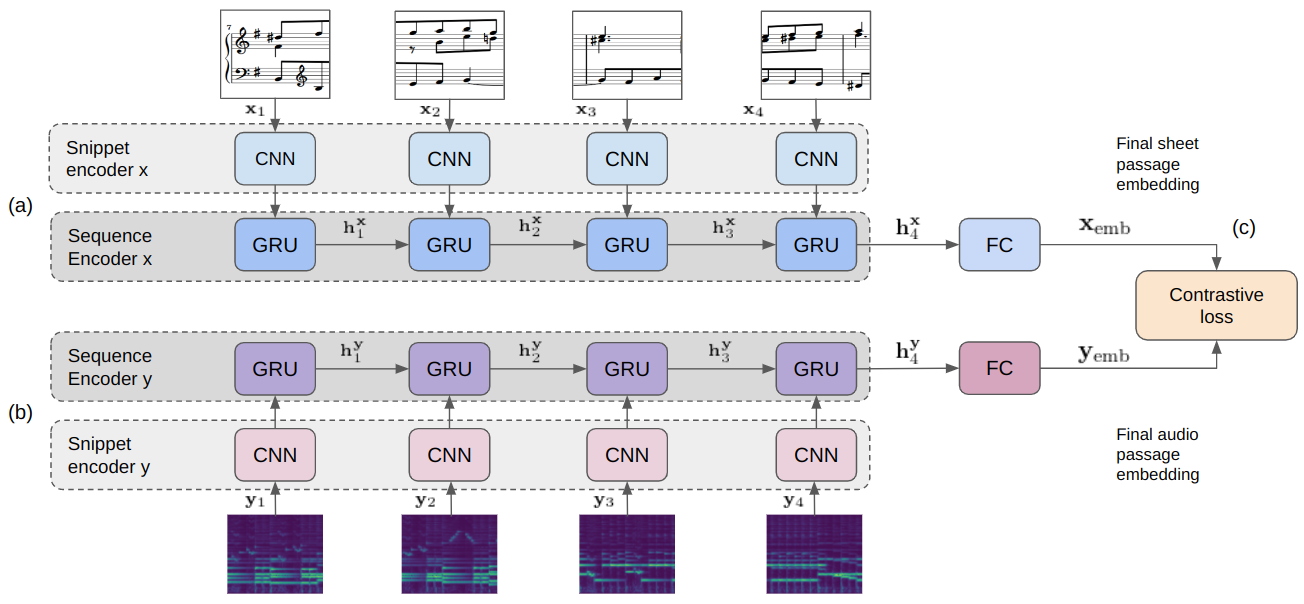}
  \caption{Diagram of the proposed network. Two independent pathways are trained 
  to encode sheet music (a) and audio (b) passages by minimizing a contrastive loss function (c).}
\label{fig:main}
\end{figure*}

A first limitation of this strategy relates to its supervised nature: 
it requires strongly-aligned data in order to generate matching audio--sheet  
snippet pairs for training, which means fine-grained mappings between note onsets and corresponding note positions in the score. 
Obtaining such annotations is tedious and time-consuming, and also requires 
specialized annotators with musical training.
As a result, embedding learning approaches have been trained with synthetic data, in
which recordings, sheet music images, and their respective alignments are rendered 
from symbolic scores.
This leads to poor generalization in scenarios with real music data, as shown 
in~\cite{CarvalhoWW23_SelfSupLearning_ASR_ACM-MMSys}.

Moreover, the snippets in both modalities have to be fixed in size, meaning that
the amount of actual musical content in the fragments can vary considerably depending 
on note durations and the tempo in which the piece is played.
For example, a sheet excerpt with longer notes played slowly would correspond to a 
considerably larger duration in audio than one with short notes and a faster tempo.
This leads to generalization problems caused by differences between what the model 
sees during training and test time; 
\cite{BalkeDCAW19_ASR_TempoInv_ISMIR} attempted to address this limitation 
by introducing a soft-attention mechanism to the network.

In this paper we address the two aforementioned limitations by proposing a 
recurrent cross-modal network that learns compact, fixed-size representations from longer 
variable-length fragments of audio and sheet music.
By removing the fixed-size fragment constraint, we can adjust the lengths of fragments 
during training so that cross-modal pairs can span the same music content, leading 
to a more robust representation.
%
Moreover, by operating with longer music passages, it is possible to 
rely solely on weakly-annotated data for training, since we now 
require only the starting and ending 
positions of longer-context music fragments within music documents, in 
order to extract audio--sheet passages to prepare a train set.
This is a remarkable advantage compared for example
to other approaches based on~\cite{DorferHAFW18_MSMD_TISMIR}, where 
fine-detailed alignments are indispensable to generate short audio--sheet  
snippet pairs.

%

The rest of the paper is structured as follows. 
In Section~\ref{sec:method} we describe the model proposed to learn joint 
representations from cross-modal passages.
Section~\ref{sec:exp} presents a series of experiments on artificial and real data 
and Section~\ref{sec:conc} summarizes and concludes the work.

\section{Audio--sheet passage retrieval}\label{sec:method}

For the purposes of this paper, and in order to be able to use our annotated corpora 
for the experiments, we define a "passage" as the musical content corresponding to 
one line of sheet music (also known as a "system"). System-level annotation of 
scores are much easier to come by than note-precise score-recording alignments, 
making it relatively easy to compile large collections of training data for 
our approach.
Our definition of passages resembles that of "musical themes", which has been 
used under a cross-modal retrieval scenario with symbolic queries in a number 
of previous works~\cite{ZalkowMueller20_WeaklyAlignedCTC_ISMIR,
BalkeALM16_BarlowRetrieval_ICASSP}.
To illustrate the temporal discrepancies between passages, we show in 
Figure~\ref{fig:histogram} the distribution of time duration of the systems 
from all pieces of the MSMD dataset~\cite{DorferHAFW18_MSMD_TISMIR}
(later we will elaborate more on this database).
In this dataset, we observe that systems can cover from less than five to 
more than 25 seconds of musical audio.


\begin{table}[t]
\centering
\scalebox{0.8}{
\begin{tabular}{cc}
\toprule
\textbf{Audio CNN encoder} & \textbf{Sheet-Image CNN encoder}\\
input: $92 \times 20$ & input: $160 \times 180$\\
\midrule
2x Conv($3$, pad-1)-$24$ - BN	    	&	2x Conv($3$, pad-1)-$24$ - BN \\
MaxPooling(2)          				&	MaxPooling(2) \\
2x Conv($3$, pad-1)-$48$ - BN			&	2x Conv($3$, pad-1)-$48$ - BN \\
MaxPooling(2)             				&	MaxPooling(2) \\
2x Conv($3$, pad-1)-$96$ - BN			&	2x Conv($3$, pad-1)-$96$ - BN \\
MaxPooling(2)          				&	MaxPooling(2) \\
2x Conv($3$, pad-1)-$96$ - BN			&	2x Conv($3$, pad-1)-$96$ - BN \\
MaxPooling(2)                    		&	MaxPooling(2) \\
Conv($1$, pad-0)-$32$ - BN 	        &   Conv($1$, pad-0)-$32$ - BN  \\
FC($32$)  	    			        &	FC($32$) \\
\bottomrule
\end{tabular}}
\caption{Overview of the two convolutinal encoders. Each side is responsible 
for their respective modality.
Conv($3$, pad-1)-$24$: 3$\times$3 convolution, 24 feature maps and zero-padding 
of 1. BN: Batch normalization~\cite{IoffeS15_BatchNorm_ICML}. We use ELU 
activation functions~\cite{ClevertUH16_ELU_ICLR} after all convolutional and 
fully-connected layers.}
\label{tab:cnn_enc}
\end{table}

This important temporal aspect motivates us to propose the network depicted in 
Figure~\ref{fig:main} to learn a common latent
representation from pairs of audio--sheet passages.
The architecture has two independent recurrent-convolutional pathways, which are 
responsible for encoding sheet music (Figure~\ref{fig:main}a) and audio 
(Figure~\ref{fig:main}b) passages.
The key component of this approach is the introduction of two recurrent layers that,
inspired by traditional sequence-to-sequence models~\cite{SutskeverVL14_Seq2Seq_NIPS}, 
are trained to summarize a variable-length sequences into context vectors, that we 
conveniently refer to as embedding vectors.

Defining a pair of corresponding passages in the form of image (sheet music) and
log-magnitude spectrogram (audio) as $\mathbf{X}$ and $\mathbf{Y}$, respectively, two 
sequences $(\mathbf{x_1}, \mathbf{x_2}, \dots, \mathbf{x_N})$ and 
$(\mathbf{y_1}, \mathbf{y_2}, \dots, \mathbf{y_M})$ are generated by sequentially cutting 
out short snippets from $\mathbf{X}$ and $\mathbf{Y}$.
The shapes of the short sheet and audio snippets are respectively $160 \times 180$ 
(pixels)\footnote{In our approach, all sheet music pages are initially re-scaled to a
$1181 \times 835$ resolution} and $92 \times 20$ (frequency bins $\times$ frames), which corresponds to one second 
of audio.
After that, each individual snippet is encoded by a VGG-style 
CNN~\cite{SimonyanZ14_VGGStyle_ICLR} into a 32-dimensional vector, as 
shown in Figure~\ref{fig:main}, generating two sequences of encoded snippets, one 
for the audio passage, and the other for the sheet passage (note that each modality 
has its own dedicated CNN encoder).
The architecture of the CNN encoders are detailed in Table~\ref{tab:cnn_enc}.

Then each sequence is fed to a recurrent layer in order to learn the spatial 
and temporal relations between subsequent snippets, which are inherent in music.
After experimenting with two typical simple recurrent layers, namely long 
short-term memory cells (LSTM)~\cite{HochSchm97} and gated recurrent units 
(GRU)~\cite{Choetal14_GRU_EMNLP}, we observed on average better results with GRUs,
and we decided for the latter for our architecture.
Each of the two GRUs is designed with 128 hidden units, 
where the hidden state of each GRU after the last step is the context vector that 
summarizes the passages.
Finally a fully connected layer (FC) is applied over each context vector, in order to encode the final passage embeddings $(\mathbf{x}_\mathrm{emb}, \mathbf{y}_\mathrm{emb})$ 
with the desired dimension.

During training, a triplet (contrastive) loss 
function~\cite{KirosSZ14_VisualSemanticEmbeddings_arxiv}
is used to minimize the distances between embeddings from corresponding passages of 
audio and sheet music and maximize the distance between non-corresponding ones.
Defining $\mathrm{d}(\cdot)$ as the cosine distance, the loss function is given by:
\begin{equation}
\mathcal{L} = 
\sum_{k=1}^K
\mathrm{max} \Bigl\{
0, \alpha + 
\mathrm{d} \Bigl( \mathbf{x}_\mathrm{emb},\mathbf{y}_\mathrm{emb} \Bigl) - 
\mathrm{d} \Bigl( \mathbf{x}_\mathrm{emb},\mathbf{y}^k_\mathrm{emb} \Bigl)
\Bigl\} \mathrm{,}
\end{equation}\label{eq:rank_loss}
where $\mathbf{y}^k_\mathrm{emb}$ for $ k \in {1, 2, \ldots, K} $ are  
contrastive (negative) examples from $K$ non-matching passages in the same 
training mini-batch.
This contrastive loss is applied to all 
$(\mathbf{x}_\mathrm{emb},\mathbf{y}_\mathrm{emb})$ pairs within each 
mini-batch iteration.
The margin parameter $\alpha \in \mathbb{R}_{+}$, in 
combination with the $\mathrm{max}\left \{ \cdot \right \}$ function, 
penalizes matching snippets that were poorly embedded.

For the sake of simplicity, we leave the remaining details concerning the design of 
the networks, such as learning hyper-parameters, to our repository where our 
method will be made publicly 
available,\footnote{\url{https://github.com/luisfvc/lcasr}} 
as well as the trained models derived in this work.

\section{Experiments}\label{sec:exp}

In this section we conduct experiments on different audio--sheet
music scenarios. 
We first elaborate on the main dataset used for training and evaluation and define 
the steps of the passage retrieval task.
Then we select four experiment setups and present the results.

We train our models with the Multi-Modal Sheet Music Dataset 
(MSMD)~\cite{DorferHAFW18_MSMD_TISMIR}, which is a 
collection of classical piano pieces with multifaceted data, including score sheets 
(PDF) engraved via Lilypond\footnote{\url{http://www.lilypond.org}} and corresponding 
audio recordings rendered from MIDI with several types of piano soundfonts.
%
%
With over 400 pieces from over 50 composers, including Bach, Beethoven and Schubert, and  
covering more than 15 hours of audio, the MSMD has 
audio--sheet music alignments which allow us to obtain corresponding cross-modal 
pairs of musical passages.
From the MSMD we were able to derive roughly 5,000 audio--sheet passages for training, which is scaled up to around 40,000 different pairs after 
data augmentation: audios are re-rendered with different soundfonts and have
their tempo changed between 90\% and 110\%.
Then we generate a test set of 534 pairs from a separate set of music pieces,
that were rendered with a soundfont that was not seen during training.
Later, in~\ref{subsec:exp_2}, we will also consider real scanned scores and real audio recordings.

To perform cross-modal passage retrieval, we first embed all audio--sheet pairs in 
the shared space using our trained model depicted in Figure~\ref{fig:main}. 
Then the retrieval is conducted by using the cosine distance and nearest-neighbor 
search within the space.
For example, in case of using an audio passage as a query to find the appropriate sheet 
music fragment, the pairwise cosine distances between the query embedding and 
all the sheet music passage embeddings are computed.
Finally, the retrieval results are obtained by means of a ranked list through sorting the distances in ascending order.

As for evaluation metrics, we look at the \textit{Recall@k} (R@k), 
\textit{Mean Reciprocal Rank} (MRR) and the \textit{Median Rank} (MR).
The R@k measures the ratio of queries which were correctly
retrieved within the top $k$ results. 
The MRR is defined as the average value of the reciprocal rank over all queries.
MR is the median position of the correct match in the ranked list.

\subsection{Experiment 1: Embedding dimension}\label{subsec:exp_1}

In the first round of experiments, we investigate the effect of 
the final embedding dimension in the retrieval task.
We consider the values in $\{16, 32, 64, 128, 256, 512, 1024\}$ and train
the model of Figure~\ref{fig:main} with the same hyperparameters.
Then we perform the retrieval task in both search directions:
audio-to-sheet music (A2S) and sheet music-to-audio (S2A).

\begin{figure}
 \centerline{
 \includegraphics[width=1\columnwidth]{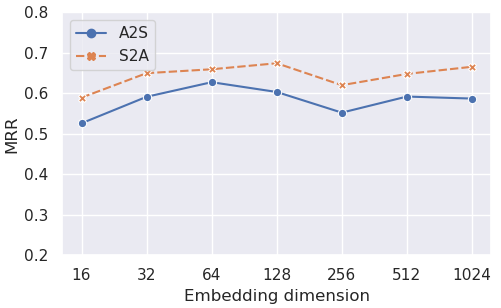}}
 \caption{Mean Reciprocal Rank (MRR) for different embedding 
 dimensions, evaluated in both search directions.}
 \label{fig:emb_sweep}
\end{figure}

\begin{table*}[!t]
\centering
  \caption{Results of audio--sheet music passage retrieval, performed in 
  both search directions, and evaluated in three types of data:
  (I) fully synthetic, (II) partially real 
  and (III) entirely real.
  Boldfaced rows represent the best performing model per dataset.}
  \label{tab:comparison}
  \scalebox{1}{
 \begin{tabular}{lccccc|ccccc}
 & \multicolumn{5}{c}{\textbf{Audio-to-Score (A2S)}} & \multicolumn{5}{c}{\textbf{Score-to-Audio (S2A)}} \\ \cmidrule{2-11}
 & \bfseries R@1 & \bfseries R@10 & \bfseries R@25 & \bfseries MRR & \bfseries MR &
   \bfseries R@1 & \bfseries R@10 & \bfseries R@25 & \bfseries MRR & \bfseries MR \\
\midrule
\midrule
\multicolumn{11}{l}{I \ \ \  MSMD (Fully synthetic)} \\
\midrule
BL         & 47.56 & 81.68 & 90.80 & 0.592 & 1    & 51.37 & 83.51 & 92.59 & 0.628 & 1 \\
RNN        & 51.12 & 84.46 & 92.88 & 0.627 & 1    & 54.30 & 85.95 & 94.94 & 0.670 & 1 \\
RNN-FT     & 55.27 & 87.98 & 95.02 & 0.651 & 1    & 56.32 & 87.12 & 96.44 & 0.697 & 1 \\
RNN-FT-CCA & \textbf{60.04} & \textbf{89.66} & \textbf{97.73} & \textbf{0.692} & \textbf{1}    & \textbf{62.11} & \textbf{91.44} & \textbf{98.41} & \textbf{0.734} & \textbf{1} \\
RNN-FZ     & 50.76 & 84.20 & 92.11 & 0.619 & 1    & 52.90 & 85.21 & 94.12 & 0.658 & 1 \\
RNN-FZ-CCA & 52.67 & 86.46 & 92.88 & 0.635 & 1    & 55.67 & 86.30 & 95.34 & 0.682 & 1 \\
\midrule
\midrule
\multicolumn{11}{l}{II \ \ RealScores\_Synth (Sheet music scans and synthetic recordings)} \\
\midrule
BL         & 20.19 & 55.47 & 74.99 & 0.343 & 7    & 25.15 & 70.27 & 83.11 & 0.391 & 5 \\ 
RNN        & 25.09 & 61.24 & 78.27 & 0.374 & 5    & 30.15 & 72.47 & 86.89 & 0.439 & 3 \\ 
RNN-FT     & 28.87 & 66.41 & 81.32 & 0.447 & 4    & 33.98 & 75.47 & 88.51 & 0.462 & 2 \\ 
RNN-FT-CCA & \textbf{33.36} & \textbf{69.49} & \textbf{83.88} & \textbf{0.481} & \textbf{3}    & \textbf{37.35} & \textbf{79.22} & \textbf{89.95} & \textbf{0.538} & \textbf{1} \\ 
RNN-FZ     & 25.83 & 62.02 & 79.74 & 0.376 & 5    & 31.45 & 74.87 & 87.26 & 0.442 & 3 \\ 
RNN-FZ-CCA & 26.82 & 63.33 & 80.19 & 0.391 & 5    & 33.55 & 75.71 & 88.79 & 0.467 & 2 \\ 
\midrule
\midrule
\multicolumn{11}{l}{III \ RealScores\_Rec (Sheet music scans and real recordings)} \\
\midrule
BL         & 15.67 & 31.46 & 48.12 & 0.226 & 29   & 18.30 & 36.71 & 54.94 & 0.266 & 18 \\
RNN        & 19.11 & 35.98 & 53.65 & 0.278 & 21   & 22.76 & 39.95 & 57.47 & 0.303 & 15 \\
RNN-FT     & 22.39 & 39.53 & 57.19 & 0.338 & 18   & 26.76 & 42.77 & 59.38 & 0.371 & 7 \\
RNN-FT-CCA & \textbf{26.62} & \textbf{44.81} & \textbf{60.01} & \textbf{0.362} & \textbf{7}   & \textbf{29.84} & \textbf{46.71} & \textbf{60.88} & \textbf{0.435} & \textbf{4} \\
RNN-FZ     & 17.65 & 33.12 & 52.98 & 0.252 & 22   & 19.13 & 37.51 & 55.57 & 0.277 & 17 \\
RNN-FZ-CCA & 18.38 & 35.81 & 54.51 & 0.279 & 21   & 22.30 & 38.95 & 58.82 & 0.285 & 16 \\
\end{tabular}}
\end{table*}

Figure~\ref{fig:emb_sweep} presents the MRR of the snippet retrieval results
evaluated on the 534 audio--sheet music passage pairs of the MSMD testset.
A first and straightforward observation is that in all cases the S2A direction 
indicates better retrieval quality.
We observe the performance increasing together with the embedding 
dimensionality until it stagnates at 64-D, and the MRR does not improve 
on average for higher-dimensional embeddings.
%
%
For this reason, we select the model that generates 64-dimensional embeddings 
as the best one, which will be evaluated more thoroughly in the next experiments.

\subsection{Experiment 2: Real data and improved models}\label{subsec:exp_2}

In this section, we conduct an extensive series of experiments comparing 
our proposed recurrent network and some improved models thereof with 
baseline methods, and extend the evaluation to real-world piano data.

%
Given that our training data are entirely synthetic,
we wish to 
investigate the generalization of our models from synthetic to real data.
To this end, we evaluate on three datasets: on a (1) fully 
artificial one, and on datasets consisting (2) partially 
and (3) entirely of real data.
For (1) we use the test split of MSMD and for (2) and (3) we 
combine the Zeilinger and Magaloff
Corpora~\cite{CancinoChaconGWG_Magaloff_ML} 
with a collection of commercial recordings 
and scanned scores that we have access to.
These data account for more than a thousand pages of sheet music scans 
with 
mappings to both MIDI files and over 20 hours 
of classical piano recordings.
Then, besides the MSMD (I), we define two additional evaluation sets:
(II) \textit{RealScores\_Synth}: a partially real set, with \textit{scanned} (real) 
scores of around 300 pieces aligned to \textit{synthesized} MIDI 
recordings.
And (III) \textit{RealScores\_Rec}: an entirely real set, with \textit{scanned} (real) 
scores of around 200 pieces and their corresponding
\textit{real audio} recordings.

As a baseline (BL), we implement the method from~\cite{DorferHAFW18_MSMD_TISMIR} 
and adapt their short-snippet-voting strategy to identify and retrieve entire
music recordings and printed scores so it can operate with 
passages.\footnote{The reasons we did not use the attention-based method 
from~\cite{BalkeDCAW19_ASR_TempoInv_ISMIR} as a baseline comparison are 
twofold. First we intend to compare the exact original snippet embedding 
architecture with and without a recurrent encoder, and adding the attention 
mechanism to a baseline model would introduce a significant number of 
additional trainable parameters, making the comparison unfair. Second, the 
purpose of the attention model is to compensate the musical content discrepancy 
between audio and sheet snippets, which is not the case for musical passages 
as defined here: 
pairs of audio--sheet music passages comprise the exact musical 
content (that is the reason why fragments are not fixed in time).}
In essence, short snippets are sequentially cut out from a passage query and 
embedded, and are compared to all embedded snippets which were selected from 
passages in a search dataset of the counterpart modality, resulting in a 
ranked list based on the 
cosine distance for each passage snippet.
Then the individual ranked lists are combined into a single ranking, in which 
the passage with most similar snippets is retrieved as the best match.

Additionally, we investigate whether our models can benefit from pre-trained 
cross-modal embeddings.
Since both CNN encoders of our proposed network architecture (see 
Figure~\ref{fig:main}) are the same as in~\cite{DorferHAFW18_MSMD_TISMIR}, 
we re-designed the baseline cross-modal network to accommodate our snippet 
dimensions ($160 \times 180$ and $92 \times 20$, for sheet and audio, respectively)
and trained a short-snippet embedding model also with the MSMD,
 as a pre-training step, and then 
loaded the two CNN encoders of our recurrent network with their respective pre-trained weights 
before training.
Our hypothesis is that, by initializing the CNN encoders with parameters that 
were optimized to project short pairs of matching audio--sheet snippets close 
together onto a common latent space, models with better embedding capacity can 
be obtained.
After loading the two CNNs with pre-trained weights, we can either freeze (FZ) them 
during training or just fine-tune (FT) on them.
Therefore, in our experiments, we refer to these modifications of our proposed vanilla
recurrent network (RNN) as RNN-FZ and RNN-FT, respectively.

Moreover, an additional CCA (canonical correlation analysis) 
layer~\cite{DorferSVKW18_CCALayer_IJMIR} is used 
in~\cite{DorferHAFW18_MSMD_TISMIR} to increase the correlation 
of corresponding pairs in the embedding space.
This CCA layer is refined in a post-training step, and we investigate 
whether this refinement
process is beneficial to our network.
In our experiments we refer to models that were initialized with pre-trained
parameters from networks that had their CCA layer refined as RNN-FZ-CCA and 
RNN-FT-CCA.

Table~\ref{tab:comparison} presents the results for all data configurations 
and models defined previously.
To keep our experiments consistent and the comparison fair, we randomly select
534 passage pairs from sets (II) and (III) to create the retrieval scenario for 
their respective experiments.

An evident observation from the table is the considerable performance 
drop as we transition from synthetic to real music data.
For all the models, the MRR drops at least 0.2 points 
to a partially real test set, and drops more than 0.3 points when moving to the 
entirely real data.
%
Moreover, as mentioned in Subsection~\ref{subsec:exp_1}, the passage retrieval 
metrics of the S2A direction are better than those of A2S for all models and 
scenarios.

Our recurrent model RNN and its variants outperform the baseline approach 
in all retrieval scenarios for all evaluation metrics.
%
%
In our findings, 
we did not see noticeable improvements when the pre-loaded encoders were 
frozen during training. 
%
In fact, for some configurations (scenarios I and III) the evaluation metrics 
were slightly worse than those from the vanilla RNN model.
When the CNN encoders are pre-loaded and enabled for fine-tuning, we observe 
the largest improvements over RNN and subsequently over BL.
Moreover, the models initialized with pre-trained weights from CCA-refined 
networks (RNN-FT-CCA) achieved the best overall results, for all test datasets 
and search directions.

In addition to the overall absolute improvements, we observe that the performance drop 
between synthetic and real datasets shrinks with our proposed models, specially
with RNN-FT-CCA.
In comparison with the baseline, the I-to-III MRR gap is reduced by 0.036 and 0.06 
points in the directions A2S and S2A, respectively.

The results we obtained and summarized in Table~\ref{tab:comparison} indicate 
that introducing a recurrent layer to learn longer contexts of musical 
content is beneficial in our cross-modal retrieval problem.
However the real-data generalization problem is still evident, and
%
in Section~\ref{sec:conc} we discuss potential solutions to address such
issues.

%

\subsection{Experiment 3: Global tempo variations}\label{subsec:exp_3}

In this experiment, we investigate the robustness of our system to global 
tempo changes.
To this end, the pieces of the MSMD test dataset are re-rendered with different
tempo ratios $\rho \in \{0.5, 0.66, 1, 1.33, 2\}$ ($\rho=0.5$ means the 
tempo was halved and $\rho=2$ stands for doubling the original tempo).
A similar study was conducted in~\cite{BalkeDCAW19_ASR_TempoInv_ISMIR} for
retrieval of short audio--sheet snippets.

Table~\ref{tab:tempo} summarizes the MRR values obtained for each tempo 
re-rendering, where the baseline method is compared with our proposed
recurrent model.
We notice the general trend that the MRR gets worse as the tempo ratio
is farther from $\rho=1$ (original tempo).
This behavior is somehow expected because the new tempo renditions are more
extreme than the tempo changes the model has seen during training.

Besides the better MRR values of the proposed network, an important improvement
concerns the performance drop when changing from $\rho=1$ to $\rho=0.5$ (slower 
renditions).
The MRR gap between these tempo ratios drops from 0.12 to 0.1 and 
from 0.09 to 0.07 points for the A2S and S2A directions, respectively, when comparing 
our network with the baseline.
This indicates that the recurrent model is more robust to global tempo variations 
and can operate well with longer audio passages.

\begin{table}[t!]
 \centering
 \begin{subtable}{\columnwidth}
 \centering
 \scalebox{0.9}{
 \begin{tabular}{lccccc}
 \toprule
 \textbf{Model} & $\rho=0.5$ & $\rho=0.66$ & \bfseries $\rho=1$ & $\rho=1.33$ & $\rho=2$ \\
 \midrule
 BL  & 0.47 & 0.54 & 0.59 & 0.52 & 0.40\\
 RNN & 0.53 & 0.59 & 0.63 & 0.58 & 0.43\\
 \bottomrule
\end{tabular}}
\caption{A2S search direction.}
\label{tab:tempo_a2s}
\end{subtable}

\begin{subtable}{\columnwidth}
 \centering
 \scalebox{0.9}{
 \begin{tabular}{lccccc}
 \toprule
 \textbf{Model} & $\rho=0.5$ & $\rho=0.66$ & \bfseries $\rho=1$ & $\rho=1.33$ & $\rho=2$ \\
 \midrule
 BL  & 0.54 & 0.59 & 0.63 & 0.56 & 0.48\\
 RNN & 0.60 & 0.64 & 0.67 & 0.61 & 0.50\\
 \bottomrule
\end{tabular}}
\caption{S2A search direction.}\label{tab:tempo_s2a}
\end{subtable}
 \caption{MRR for different tempo renderings of the test pieces of MSMD 
 in both (a) audio-to-sheet and 
 (b) sheet-to-audio retrieval directions.
 We evaluate both baseline and RNN models.}
\label{tab:tempo}
\end{table}

\subsection{Experiment 4: Qualitative analysis}\label{subsec:exp_4}

\begin{figure}[t!]
 \centerline{
 \includegraphics[width=1\columnwidth]{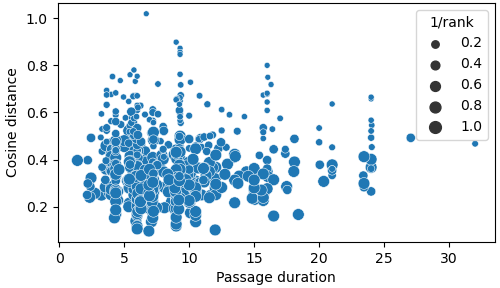}}
 \caption{Cosine distance in the embedding space in relation to the respective 
 audio passage duration of 534 pairs from the MSMD test set. The cosine distances
 were computed with the RNN model.}
 \label{fig:scatter}
\end{figure}

To get a better understanding of the behavior of our proposed network, 
in this last experiment we take a closer look at the shared embedding space 
properties. 
Figure~\ref{fig:scatter} shows the distribution of the pairwise cosine distances 
between the passage pairs from the MSMD test set, in relation to the duration (in 
seconds) of their respective audio passages.
Moreover, we scale the point sizes in the plot so they are proportional to their
individual precision values (inverse of the rank values), when considering the S2A 
experimental setup.

An interesting behavior in this visualization is the size of the points increasing 
as the cosine distance decreases.
It is expected that passage pairs with smaller distances between them,
meaning that they are closer together in the embedding space, would be 
lead to better retrieval ranks. 

Another interesting aspect of this distribution concerns the proportion
of larger cosine distances as the audio duration of the passages increases.
For example, between five and ten seconds, there are more large points 
observed than smaller ones, while between 20 and 25 seconds, the proportion
is roughly equal.
This indicates that, in our test set, embeddings from shorter passages of 
audio are still located closer to their sheet counterparts in comparison with
longer audio passages, despite our efforts to design a recurrent networks that 
learns from longer temporal contexts.

\section{Conclusion and future work}\label{sec:conc}

We have presented a novel cross-modal recurrent network 
for learning correspondences between audio and sheet music passages.
Besides requiring only weakly-aligned music data for training, this 
approach overcomes the problems of intrinsic global and local tempo mismatches 
of previous works that operate on short and fixed-size fragments.
Our proposed models were validated in a series of experiments under 
different retrieval scenarios and 
generated better results when comparing with baseline methods, for all 
possible configurations.

On the other hand, a serious generalization gap to real music data was 
observed, which points us to the next stages of our research.
A natural step towards making deep-learning-based cross-modal audio--sheet
music retrieval more robust would be to include real and diverse data that can 
be used for training models.
However such data with suitable annotations are scarce, and recent advances in 
end-to-end full-page optical music recognition~\cite{RiosIC22_FullPageOMR_ISMIR} 
can be a possible solution to learn correspondences on the score page level.
Moreover, the powerful transformers~\cite{Ashish17_Attention_NIPS} are 
potential architectures to learn 
correspondences from even longer audio recordings, accommodating typical 
structural differences between audio and sheet music, such as jumps and repetitions.

\section{Acknowledgments}

This work is supported by the European Research Council
(ERC) under the EU’s Horizon 2020 research and innovation
programme, grant agreement No.~101019375 (\textit{Whither Music?}), and the Federal State of Upper Austria (LIT AI Lab).

\bibliography{ISMIR2023_template}

\end{document}